\documentclass[twocolumns,11pt]{elsarticle}
\usepackage[a4paper, total={6in, 8in}]{geometry}
\usepackage{amssymb}
\usepackage{tabularx} 
\usepackage{amsmath}  
\usepackage{graphicx} 
\usepackage[dvipsnames]{xcolor} 
\usepackage[final]{hyperref} 
\hypersetup{
	colorlinks=true,       
	linkcolor=blue,        
	citecolor=blue,        
	filecolor=magenta,     
	urlcolor=blue
}
\usepackage{float}
\usepackage{lineno}
\usepackage{soul}
\journal{Fusion Engineering and Design}

\begin{document}

\begin{frontmatter}

\title{Relativistic effects on stopping power of plasmas with heavy ions}

\author[UCLM,CONICET]{A. M. P. Mendez}

\affiliation[UCLM]{
    organization={E.T.S.I. Industrial, Universidad de Castilla-La Mancha},
    city={Ciudad Real},
    postcode={13071}, 
    country={Spain}
}

\affiliation[CONICET]{
    organization={IAFE, CONICET and Universidad de Buenos Aires},
    city={Buenos Aires},
    postcode={C1428EGA}, 
    country={Argentina}
}

\author[UCLM]{J. Chacón-Gijón}
\author[UCLM]{J. Vázquez-Moyano}
\author[UCLM]{M. D. Barriga-Carrasco}

\begin{abstract}

This work investigates the electronic stopping power for protons of a tungsten plasma at various electron densities and temperatures. The study employs a dielectric formalism to model the stopping power due to free and bound electrons, considering relativistic and non-relativistic effects. The bound electron stopping power is modeled using the shellwise local plasma approximation (SLPA) introduced by Montanari and Miraglia. The ionization state of the plasma is also examined, revealing its impact on the bound electron contribution. 
We compare our results with the T-Matrix approach and the Li-Petrasso model for free electrons in combination with SLPA calculations for the bound electron ones. Results demonstrate the significance of bound electron stopping power, particularly for plasma with high-$Z$ ions. This investigation contributes valuable insights into plasma physics and fusion energy research, providing essential data for future experiments and simulations.

\end{abstract}


\end{frontmatter}

\section{Introduction}

There are two main approaches to fusion energy: magnetic confinement fusion (MCF) and inertial confinement fusion (ICF). Due to its properties, tungsten is a plasma-facing material candidate, which will be directly exposed to the target explosion. Several works have investigated the possibility of using tungsten lines to allow plasma diagnosis using spectroscopy~\cite{Clementson2010,Murakami2021,Gao2022}. Other studies have examined the capability of high-$Z$ dopants as stabilizers of the fuel~\cite{Hopkins2018,Li2021}. Because the presence of tungsten in fusion plasmas is indisputable, the characterization of the energy loss processes for ions in plasma with tungsten is vital. 

During laser-driven ICF implosion experiments, laser pulses ablate the surface of a layered target and implode the target~\cite{Nuckolls1972}. The laser-driven shocks and the spherical convergence compress the fuel, creating extreme temperatures and pressures. The high temperatures and densities cause the fuel to fuse, producing energetic $\alpha$ particles and neutrons. 
The implosion generates a relatively low-temperature and high-density shell and a high-temperature and medium-density hotspot center. The $\alpha$ particles created from the reaction in the target center carry the necessary energy to heat the plasma further. As the particles travel through the plasma, they deposit their energies. The energy loss per unit length, also called stopping power, is directly involved in determining the effectiveness of this process. 

In this work, we aim to study the electronic stopping power for protons of tungsten plasmas at various electron densities and temperatures. Although heavy-ion plasma measurements have gone as heavy as Xe~\cite{Mintsev1999, BarrigaLPB13}, we theoretically examine a very heavy-ion plasma such as tungsten, which features a nuclear charge $Z=74$. Studying the stopping power of such plasma may be vital for understanding the ion energy deposition and further ignition of W-doped ICF plasmas.

The electronic stopping power model and calculations are obtained considering the mean energy loss per unit length contribution of free and bound electrons separately. For the free electrons, we use a dielectric model for electron quantum plasmas in thermal equilibrium that introduces a temperature dependence. Because elements ionize with increasing temperatures, we computed the stopping power for protons due to the bound electrons of all the tungsten ions that should form in the plasma within a particular temperature range. For this, we also implemented the dielectric formalism with a dielectric function that relies on electronic structure values. We computed the electronic structure of the tungsten ions using two frameworks, a relativistic and a non-relativistic approach; then, we examined the effects of employing these models in the final stopping power results. We also compare our calculations with values from various theoretical models for each type of electron contribution.

\section{Stopping power of free electrons}
\label{sec:free}

We employ the dielectric formalism to model the stopping power for protons due to the free electrons of a tungsten plasma. For a bare ion of charge $Z_P$ moving at velocity $v$, the stopping power due to the free electrons is given by
\begin{equation}
S=\frac{2}{\pi} \frac{Z_P^2 e^{2}}{v^{2}} \int_{0}^{\infty} \frac{d k}{k} \int_{0}^{k v} d \omega \,\omega \operatorname{Im}\left[\frac{-1}{\varepsilon(k, \omega)}\right] .
\end{equation}
where $\epsilon(k,\omega) = \epsilon_1(k,\omega) + i \epsilon_2(k,\omega)$ is the dielectric function proposed by Arista and Brandt~\cite{Arista1984}.
The real and imaginary parts of this function are, respectively,
\begin{equation}
  \epsilon_{1}(k, \omega)=1+\frac{\chi_{0}^{2}}{4 z^{3}}[g(u+z)-g(u-z)],
  \label{eq:Arista_e1}
\end{equation}
with
\begin{equation}
    g(x)=-g(-x)=\int_{0}^{\infty} \frac{y d y}{e^{D y^{2}-\eta}+1} \ln \left|\frac{x+y}{x-y}\right| \,,
\end{equation}
and
\begin{equation}
\epsilon_{2}(k, \omega)=\frac{\pi \chi_{0}^{2}}{8 z^{3}} \theta \ln \left(\frac{1+\exp \left[\eta-D(u-z)^{2}\right]}{1+\exp \left[\eta-D(u+z)^{2}\right]}\right),
  \label{eq:Arista_e2}
\end{equation}
where $\chi_0^2 = 1/(\pi k_F a_0)$, $\theta = k_B T/E_F$, $D=1/\theta$, and $\eta = \beta\mu = \mu/(k_BT)$. The usual reduced variables are $u=\omega / (k v_{F})$ and $z=k / (2 k_{F})$, with $E_F = \hbar^2k_F^2/(2m)$ being the Fermi energy, and $k_{F}$ the corresponding wave vector, $a_0$ is the Bohr radius, and $k_B$ is Boltzmann's constant.

This research group has proposed other dielectric functions for free electrons in the past ~\cite{BarrigaLPB06, BarrigaLPB08, BarrigaLPB10}, but they are much more complex to deal with and have similar results to the Arista dielectric function in the present calculations.

\section{Stopping power of bound electrons}
\label{sec:bound}

The contribution of the bound electrons of the tungsten ions to the stopping power is also modeled with a dielectric function approach: the shellwise local plasma approximation (SLPA) introduced by Montanari and Miraglia~\cite{Montanari2013a}.
The SLPA formulation for the stopping power cross-section by a bare ion of charge $Z_P$ moving at velocity $v$ in the atomic cloud of $nl$ subshell electrons is given by
\begin{equation}
S_{n l}=\frac{2}{\pi} \frac{Z_P^2 e^2}{v^2} \int_{0}^{\infty} \frac{d k}{k} \int_{0}^{k v} \omega \operatorname{Im}\left[\frac{-1}{\varepsilon_{n l}(k, \omega)}\right] d \omega\,.
\end{equation}
Within the SLPA, the imaginary part of the inverse dielectric function is expressed as
\begin{equation}
\operatorname{Im}\left[\frac{-1}{\varepsilon_{n l}(k, \omega)}\right]= \int \operatorname{Im}\left[\frac{-1}{\varepsilon^{\mathrm{SLPA}}\left(k, \omega, \rho_{n l}(r), E_{nl}\right)}\right] \mathbf{d r},
\end{equation}
where $\rho_{nl}$ is the orbital radial density and $E_{nl}$ is the binding energy of the $nl$ subshell.
In this work, we used the Levine-Lindhard dielectric function~\cite{Levine1982}, i.e., $\varepsilon^{\mathrm{SLPA}}=$ $\varepsilon^{\mathrm{LL}}\left(k, \omega, \rho_{nl}, E_{n l}\right)$, which was initially proposed to describe ionization threshold or excitation gaps.
Levine and Louie~\cite{Levine1982} included explicitly the energy gap $E_{n l}$ within the Lindhard dielectric function as follows:
\begin{equation}
\operatorname{Im}\left[\frac{-1}{\varepsilon^{\mathrm{LL}}\left(k, \omega, E_{n l}\right)}\right]=\operatorname{Im}\left[\frac{-1}{\varepsilon^{\mathrm{L}}\left(k, \omega_{g}\right)}\right] \Theta\left(\omega-\left|E_{n l}\right|\right)\,,
\label{eq:SLPA-ELF}
\end{equation}
where $\varepsilon^{\mathrm{L}}(q, \omega)$ is the Lindhard dielectric function~\cite{Lindhard1954}, $\Theta(x)$ is the Heaviside step function, and $\omega_{g}=\sqrt{\omega^{2}+E_{n l}^{2}}$.
Following Lindhard and Scharff~\cite{Lindhard1953}, in the formulation proposed by Montanari and Miraglia~\cite{Montanari2013a}, the orbital radial density is introduced in the dielectric function as a local electron density with a plasmon frequency given by $\omega_{p}^{nl}(r)=\sqrt{4 \pi \rho_{nl}(r)}$.

\section{Results}

This section presents our results for the stopping power of a tungsten plasma for protons. We analyze plasmas with two electron densities $n_{fe} = 1\times10^{22}$~cm$^{-3}$ and $1\times10^{23}$~cm$^{-3}$ at three temperatures: $T=$~10~eV, 30~eV, and 60~eV. 
We obtained the present results by assuming independent free and bound electron contributions to the stopping power. 
The free electron component is modeled with the temperature-dependent dielectric formalism detailed in Section~\ref{sec:free}. 
The corresponding bound electron contribution to the stopping power is also modeled within the linear response formalism, which includes electronic structure data of the ions. Because we consider a plasma composed of heavy atoms, we should consider including relativistic corrections for describing the most inner-bound electrons. We compare our results with non-relativistic ones to account for these corrections within the same formalism.

\subsection{Temperature dependence of the ionization state of the plasma}
\label{subsec:flychk}

When the temperature in the plasma increases, tungsten ionizes; the ionization state of the target changes, and correspondingly, the fraction of free and bound electrons varies. To account for this phenomenon, we considered the \textsc{flychk} code~\cite{Chung2005}, which allows us to compute the temperature dependence of the mean ionization state $\bar{q}$ of the ion at a given electron density. 
The results obtained in this section are later used to compute the bound and free-electron stopping powers.

\begin{figure}[t]
    \centering
    \includegraphics[height=0.35\textheight]{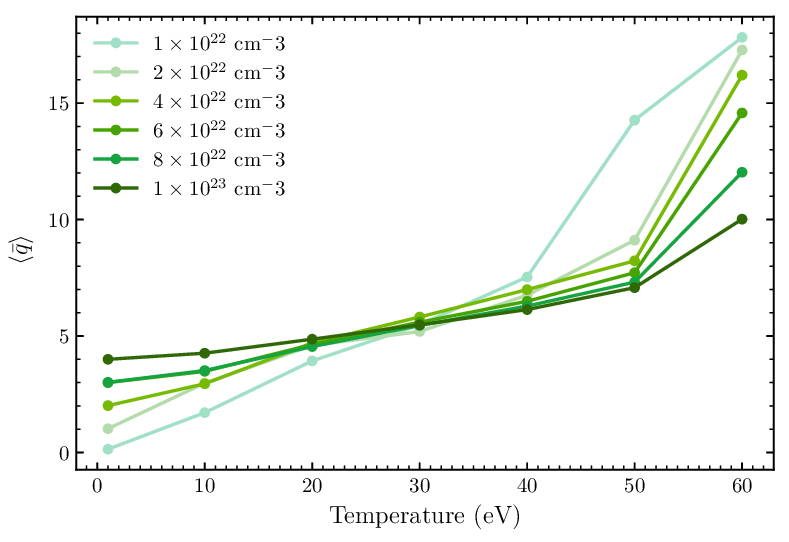}
    \caption{(color online)
    Mean ionization charge of the plasma composed of tungsten in terms of the temperature for electron densities ranging from $1\times10^{22}$~cm$^{-3}$ to $1\times10^{23}$~cm$^{-3}$.}
    \label{fig:flychk}
\end{figure}
The temperature dependence of the ionization state of the ions composing the plasma is shown in Fig.~\ref{fig:flychk} for electron densities ranging from $1\times10^{22}$~cm$^{-3}$ to $1\times10^{23}$~cm$^{-3}$. As seen in the figure, the mean charge state of the ion for the lowest electron density at 1 eV is close to 0. In contrast, this value is close to 4 for the highest electron density, i.e., for a plasma with high density, the plasma rapidly loses its most weakly bound electrons. The $\bar{q}$ values at $T=$~10 eV range from 1.7 to 4.3, while \mbox{for 30 eV}, the mean charge is close to 5.5.  In a plasma at \mbox{60 eV}, the results for $\bar{q}$ spread broadly from 10 for $n_{fe} = 1\times10^{23}$~cm$^{-3}$ to 17.8 for $n_{fe} = 1\times10^{22}$~cm$^{-3}$. In this temperature value, the dependence of the ionization state with $T$ has been reversed, i.e., in a plasma with higher electron density, the ions require more energy to lose bound electrons.

\subsection{Stopping power of free electrons}
\label{subsec:free}

\begin{figure}[t]
    \centering
    \includegraphics[height=0.35\textheight]{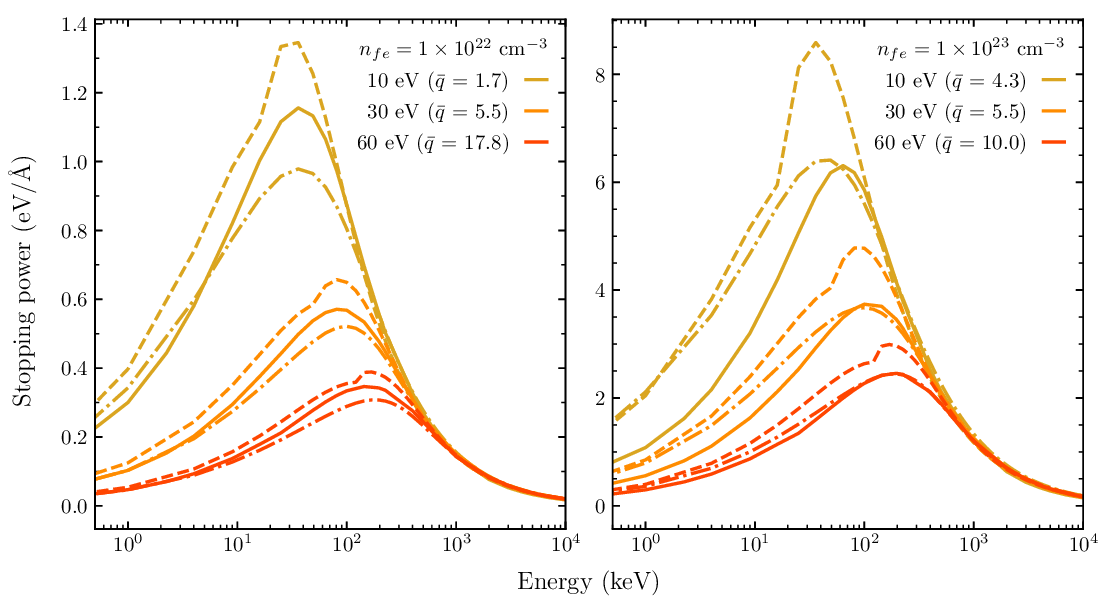}
    \caption{(color online) Free electron stopping power for protons of a plasma with $n_{fe}$ electron density at 10~eV, 30~eV, and 60~eV: present results (solid lines), Li-Petraso model (dashed lines), and T-Matrix (dash-dotted lines).
    }
    \label{fig:free-stopping}
\end{figure}
We implemented the random phase approximation with the Arista dielectric function (ADF) given by Eqs.~(\ref{eq:Arista_e1}) and (\ref{eq:Arista_e2}) to model the free electron stopping power for protons of a plasma. 
Our results are illustrated in Fig.~\ref{fig:free-stopping} with solid lines for the cases discussed before, i.e., plasma with electron densities \mbox{1$\times10^{22}$~cm$^{-3}$} (left panel) and \mbox{1$\times10^{23}$~cm$^{-3}$} (right panel) at temperatures of 10~eV, 30~eV, and 60~eV. We compare our calculations with two widely used methods: the Li-Petrasso model~\cite{Li1993} and the T-Matrix approach~\cite{Gericke1999}. The Li-Petrasso stopping power model is a perturbative model that considers short-range plasma interactions (collisions). This approximation is valid for the entire range of projectile-electron impact parameters. The T-Matrix method relies on a binary collision approach in the general kinetic equation, which describes short-range interactions between the projectile and the plasma electrons, including strong collisions.
Our results for each of the nine cases agree quite well with both theories, with our stopping maxima calculations lying between or close to one of them. 
The low energy stopping power is generally underestimated for the 10~eV and 30~eV plasma with \mbox{$n_{fe}=1\times10^{23}$~cm$^{-3}$}. These differences arise since the dielectric formalism relies on a perturbative approach, which requires the velocity of the incident projectile to be much larger than the velocity of the electrons.
Moreover, the stopping power of the plasma decreases as the temperature increases. Given that the thermal velocity of these electrons increases, the energy transfer from the projectile to electrons is more complicated.

\subsection{Stopping power of bound electrons}

We implemented the shellwise local plasma approximation (SLPA) to compute the bound electron contribution to the stopping power for protons of plasmas at various temperatures and densities. As established in Section~\ref{sec:bound}, the SLPA also relies on the dielectric formalism, featuring a dielectric function that depends on the electronic structure of the target, i.e., the orbital radial densities and binding energies. Therefore, we must compute these values for all tungsten ions with ionization states from 0 to +19, as suggested by the \textsc{flychk} results illustrated in Fig.~\ref{fig:flychk}. 

\subsubsection{Atomic structure of tungsten ions}

We are examining ions with $Z=74$; therefore, we include relativistic effects for correctly describing the electronic structure of the most inner electrons. We will compare the present relativistic calculations with non-relativistic ones to later analyze the influence of these effects on the bound electrons and total stopping power results.

We considered two self-consistent electronic structure theories: the Hartree-Fock and the Dirac-Fock methods, which are non-relativistic (NR) and relativistic (R) approaches. We employed the \textsc{hf} code by C. Fischer~\cite{FroeseFischer1987} and the \textsc{fac} code by M. Gu~\cite{Gu2011}, respectively, to implement these techniques.

\begin{figure}[t!]
    \centering
    \includegraphics[width=0.7\textwidth]{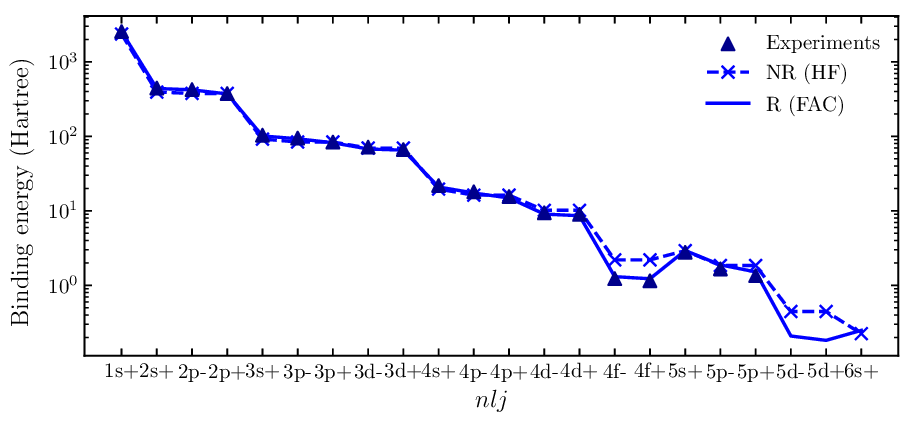}
    \caption{Binding energies of neutral tungsten: R, relativistic calculations  (solid line); NR, non-relativistic values (dashed line and $\times$ symbols); experiments in solids ($\blacktriangle$ symbols).
 }
    \label{fig:bindener-WI-RvsNR}
\vspace{1cm}
    \includegraphics[width=0.7\textwidth]{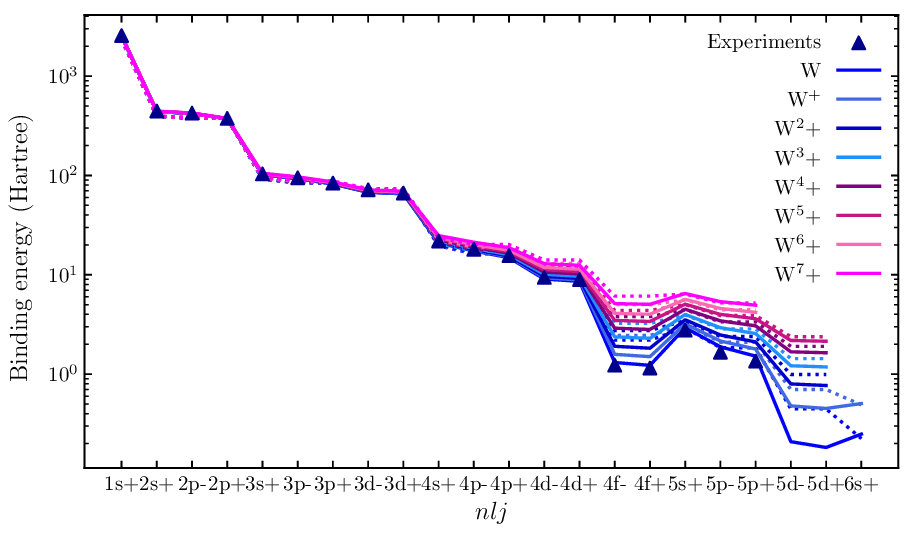}
    \caption{(color online) Relativistic (solid lines) and non-relativistic binding (dotted lines) energies of tungsten ions from neutral to W$^{7+}$. Experiments in solid tungsten are given for comparison.}
    \label{fig:binding-ener}
\end{figure}

First, we examine the orbital binding energy results for neutral tungsten. These results are illustrated in Fig.~\ref{fig:bindener-WI-RvsNR}. The non-relativistic and relativistic values are compared with measurements taken in solid tungsten. 
As expected, the inner-shell electrons are bound tighter in the R theory than in the NR one, with differences of around 10\%. Although these differences are significant, the contribution of these electrons to the stopping power will later be seen to be negligible. Therefore, the corrections to the binding energies of the inner shells are not crucial for the total stopping power calculation.
However, while relativistic corrections are included for describing the most inner shells, these effects affect the electrons of the outer shells more noticeably. The NR results agree with experimental measurements in a solid within 45\%, while these differences drop to about 8\% for the R calculations.
The improvement of the most-outer electrons (sub-valence) description is a by-product of considering the relativistic corrections in the Hamiltonian, most likely carried out by the orthonormalization condition of the orbitals. It is essential to acknowledge these effects since outer shell electrons are the ones that, as will be seen later, contribute the most to the stopping power.

To examine the effects of including relativistic corrections in the binding energies of tungsten ions, we present the binding energies of the first eight elements of its isonuclear sequence in Fig.~\ref{fig:binding-ener}. The binding energies of the most inner electrons slowly vary as the isonuclear sequence grows, with the electrons becoming more bound as the atom gets ionized. However, the behavior of valence and subvalence electrons ($4f$, $5d$, and $6s$) changes significantly as the ionization of tungsten increases. The differences between the relativistic and non-relativistic binding energies range from 3\% to 10\% and 9\% to 140\% for the $6s$ and $5d$ shell, respectively. For the $4f$ electrons, these discrepancies vary from 22\% to 80\% for the first six elements of the isonuclear sequence and from 5\% to 21\% in the remaining cases. These differences will become relevant in the stopping power results presented in the following section.

\subsubsection{Stopping power}
\label{subsubsec:bound}

The atomic structure calculations in the previous section present $nlj$ splitting, and one may be prone to consider the stopping power due to each group of spin-oriented electrons within $nl$ quantum numbers separately. However, for each subshell, the collisional time can be estimated as $\Delta t \approx \langle r \rangle_{nlj}/v$, where $\langle r \rangle_{nlj}$ is the mean radius of orbital $nlj$ and $v$ is the velocity of the incident proton. Then, considering the Heisenberg principle, the $nlj$ split is unresolved in the energy region where the electrons contribute to the stopping power and the $nl$ electrons must be considered together (\textit{intra-shell screening}). 
The same effect is considered for shells with similar binding energy and parity (\textit{inter-shell screening}). This approach has been successfully implemented to describe the stopping power of protons colliding on many heavy solid targets, such as Hf, W, Au, Pb, and Bi~\cite{Montanari2020, Mendez2022, Montanari2009a, Montanari2009b, Montanari2011}.

\begin{figure}[t!]
    \centering
    \includegraphics[height=0.35\textheight]{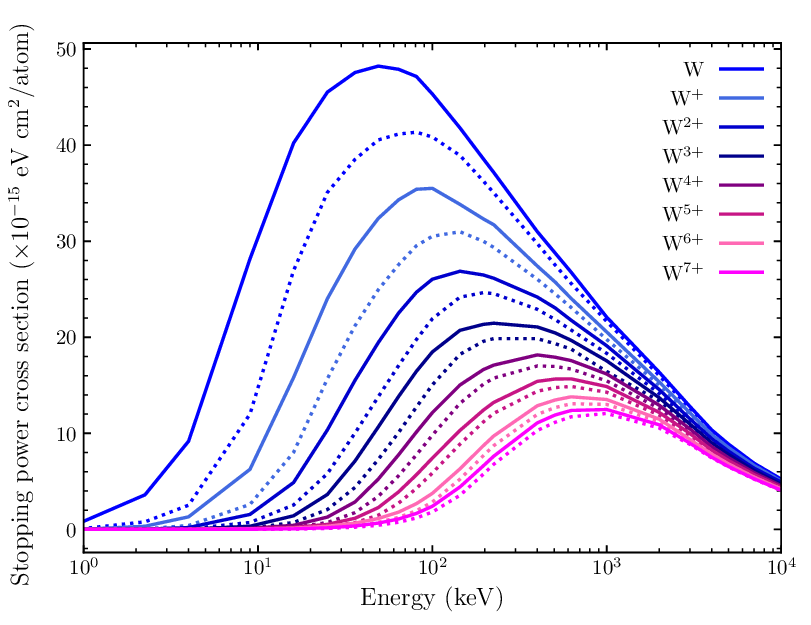}
    \caption{(color online)
    SLPA stopping power for protons of the first eight elements of tungsten's isonuclear sequence. Relativistic (solid lines) and non-relativistic (dotted lines) results include inter-shell screening of the $4f$-$5p$ and $5d$-$6s$ electrons. 
    }
    \label{fig:SLPA-isonuclear-tungsten}
\end{figure}

To determine the bound electron contribution to the stopping power for protons when interacting with tungsten ions of mean ionization charge $\bar{q}$, we first determined the stopping power due to bound electrons of the first 20 elements of the isonuclear sequence of tungsten. In Fig.~\ref{fig:SLPA-isonuclear-tungsten}, we present the results for the first eight ions. These values include the inter-shell and intra-shell shell screening for the relativistic (solid lines) and non-relativistic (dotted lines) calculations. These results correspond to implementing the SLPA with the relativistic and non-relativistic electronic structure data, i.e., $\rho_{nl}$ and $E_{nl}$, presented in the previous section. For the first ions (W I, II, and III), the stopping maxima for the relativistic calculations are larger than the non-relativistic ones by about 15\%. Furthermore, the proton energies corresponding to these stopping maxima are shifted by around 33\% to lower values as these effects are considered. However, as the ionization increases, the differences in the amplitude of the stopping cross sections are reduced to about 5\% without maxima shifts.
Remarkably, the amplitude of the stopping power due to bound electrons decreases by more than half when the ion loses only three electrons; the number of electrons in the atom goes from 74 to 71, yet the stopping maximum decreases by about 55\%. We obtain this result because tungsten loses its weakly bound electrons; these electrons are far from the screened nuclei and strongly interact with the incoming projectile. Then, as the ions lose their outer electrons, only the most strongly bound remain, which do not interact with the projectile as much. The stopping power due to remaining electrons becomes less than 10\% than in the case of neutral tungsten when the ion loses only a couple of ten electrons.

As established in Section~\ref{subsec:flychk}, the mean ionization charge of tungsten varies according to temperature and density, and the $\bar{q}$ values were shown to be fractional. We performed a cubic spline interpolation between the stopping power results of the closest ions to obtain the  energy loss per unit length due to the $Z_T-\bar{q}$ bound electrons. Indeed, the bound electron density depends on the plasma temperature through $\bar{q}$ and is computed as \mbox{$n_{be} = (Z_T - \bar{q}) \, n_{fe} / \bar{q}$}. For a plasma with density \mbox{$n_{fe} = 1\times10^{22}$~cm$^{-3}$} at $T=10$~eV, the mean charge state of tungsten is $\bar{q}=1.7$. To compute the corresponding bound electron stopping power, we utilized the results shown in Fig.~\ref{fig:SLPA-isonuclear-tungsten} for W$^{+}$ and W$^{2+}$, with a bound electron density of \mbox{$n_{be} = 4.2\times10^{23}$~cm$^{-3}$}.

\begin{figure}[t]
    \centering
    \includegraphics[height=0.35\textheight]{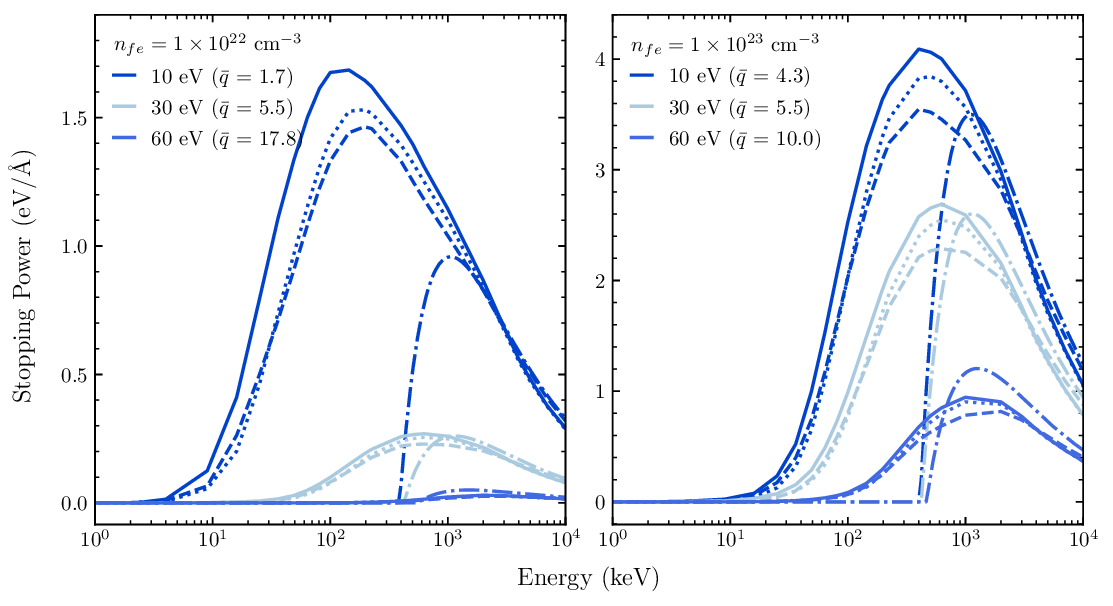}
    \caption{(color online) Bound electron stopping power for protons of plasmas with $n_{fe}$ electron density at 10~eV, 30~eV, and 60~eV: relativistic (R, solid lines) and non-relativistic (NR, dotted lines) SLPA results, EWPM (dashed lines), and Zimmerman model (dash-dotted lines).
    }
    \label{fig:bound-stopping}
\end{figure}
Finally, the present results for the bound electron contribution to the stopping power for a tungsten plasma with electron densities \mbox{1$\times10^{22}$~cm$^{-3}$} (left panel) and \mbox{1$\times10^{23}$~cm$^{-3}$} (right panel) at temperatures $T=$~10~eV, 30~eV, and 60~eV are shown in Fig.~\ref{fig:bound-stopping}. The relativistic (solid lines) and non-relativistic (dashed lines) values also decrease as the temperature increases. However, in contrast to the free electron stopping results, the bound electron stopping power varies differently as the plasma temperature increases with constant electron density. The stopping power due to bound electrons of a plasma with \mbox{$n_{fe}=1\times10^{22}$~cm$^{-3}$} decreases more rapidly as the plasma temperature increases than in the case of a plasma with \mbox{$n_{fe}=1\times10^{23}$~cm$^{-3}$}. As previously discussed, these differences are related to the mean ionization state $\bar{q}$ of the ion predominant at a given temperature. 

The SLPA results in Fig.~\ref{fig:bound-stopping} are compared with values from two independent models: the extended wave-package model (EWMP) proposed by Archubi \& Arista~\cite{Archubi2017} and the Zimmerman model~\cite{Zimmerman1997}. The EWPM is a wave-packet approach with the same binding energy threshold criteria~\cite{Levine-Louie} as the SLPA does. The Zimmerman model uses a Bethe-like expression with a mean ionization potential expression for bound electrons. The EWPM calculations have been performed with non-relativistic binding energies, momentum parameters for neutral tungsten, and without intra-shell screening. 
In most cases, our non-relativistic calculations and EWPM values agree well, with stopping power predictions below the relativistic ones. As expected, all the results agree with the Zimmerman model for proton impact energy greater than 100~keV.

\subsection{Total stopping power}

\begin{figure}[t!]
    \centering
    \includegraphics[height=0.35\textheight]{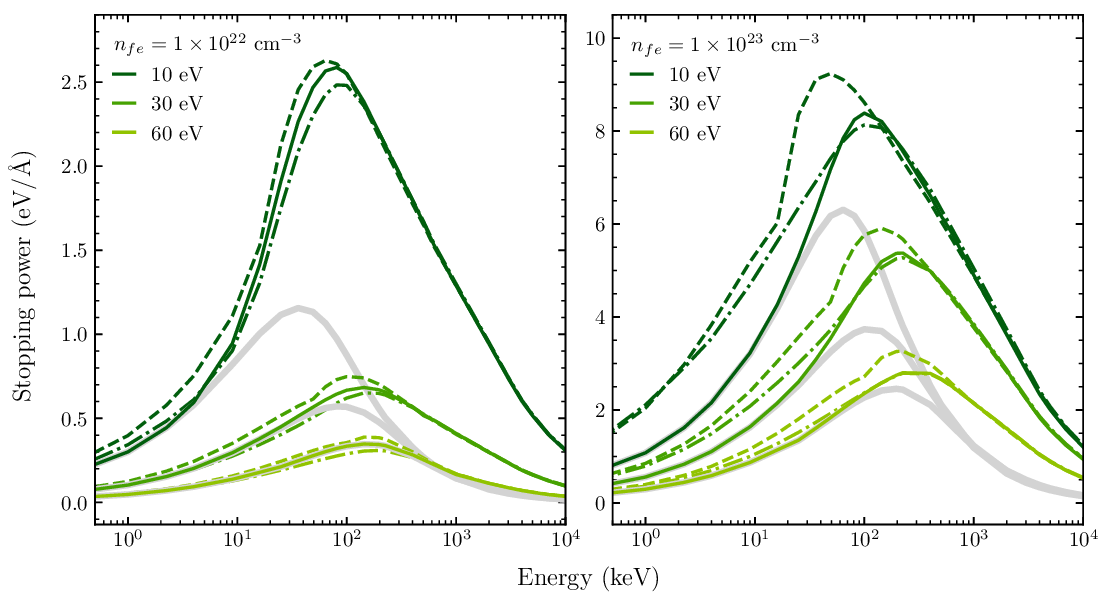}
    \caption{(color online) Total stopping power for protons of a plasma with electron density $n_{fe}$ at 10~eV, 30~eV, and 60~eV. The free electron stopping models of Fig.~\ref{fig:free-stopping} are combined with relativistic SLPA results: SLPA + ADF (solid lines), SLPA + Li-Petraso model (dashed lines), and SLPA + T-Matrix (dash-dotted lines). We included the free electron stopping power results of the Arista dielectric function formalism (thick solid lines) for comparison.}
    \label{fig:total-stopping}
\end{figure}

We combine the free electron and the bound electron contribution to obtain the total electronic stopping power for protons of a plasma. We illustrate our results in Fig.~\ref{fig:total-stopping} for all the electron densities and temperatures examined thus far. We compare the present results with the T-Matrix approach and the Li-Petrasso model for free electrons in combination with the SLPA for bound electrons. 
We also included the free electron ADF values to discuss the importance of considering the bound electron stopping power in these calculations. 

As the temperature increases, our total stopping power results in Fig.~\ref{fig:total-stopping} decrease, following the same trend as the free and bound electron contributions discussed in Sections~\ref{subsec:free} and \ref{subsubsec:bound}.
The free electrons contribute entirely to the total stopping power for slow protons at all temperatures and densities. For a plasma with electron density \mbox{$n_{fe} = 1\times10^{22}$~cm$^{-3}$}, the bound electron contribution to the total stopping power becomes relevant only for incident proton energies greater than 5~keV and 40~keV for plasmas with temperatures of 10~eV and 30~eV, respectively. For the lowest-temperature plasma, the bound electrons constitute nearly 70\% of the total stopping power around the maxima. However, for a plasma at 60~eV, the stopping power is almost entirely due to the free electrons. 
For a plasma with a free electron density of \mbox{$n_{fe} = 1\times10^{23}$~cm$^{-3}$}, the bound electron contribution becomes relevant at higher proton energies, i.e., 20~keV, 30~keV, and 80~keV for the respective increasing plasma temperatures considered. The relative contribution to the total stopping in this case is minor compared to the previously discussed case, constituting only around 30\% of the total stopping at the maxima. In this scenario, the ion has been ionized four times, losing its most weakly bound electrons. The mean charge of the ion in this plasma slowly varies in this temperature range, going from 4.3 to 10, and so does the bound electron stopping power; however, as the plasma features high bound electron densities, the stopping power due to these electrons is not negligible even at temperatures of 60~eV.

%
%

\section{Summary}

This work presents relativistic results for the electronic stopping power for protons of a tungsten plasma at temperatures $T=10$~eV, $30$~eV and $60$~eV and free electron densities $n_{fe}=1\times10^{22}$~cm$^{-3}$ and $1\times10^{23}$~cm$^{-3}$. We obtained total values by combining stopping power calculations due to free and bound electrons. The interaction of incident projectiles and free and bound electrons was modeled within the dielectric formalism, which includes collective and binary collisions. 
This approach relies on the perturbative approximation, and it is valid for incident projectile velocities much larger than the mean velocity of the electrons interacting with the projectile and for projectiles with smaller nuclear charges than the target. 

The free electron interaction with the projectiles was modeled with a quantum plasma dielectric function~\cite{Arista1984}. This approach allows one to model the stopping power of a plasma at any degeneracy. 
The energy loss per unit length for the projectile due to the bound electrons was modeled with the shellwise local plasma approximation~\cite{Montanari2013a}.
The present free and bound electron stopping power results were compared with other models. In most cases, all the theoretical models agree.

We found the bound electron contribution to stopping power can sometimes be significant, if not determinant, in high-$Z$ plasmas. For certain electron densities and temperature combinations, the stopping power of the plasma near the maxima can dominated by the bound electrons, or it can be negliglible. In other cases, the bound electron stopping power can be significant across the entire temperature range. As a final mark, bound electron contributions to the stopping power of plasmas must be addressed in cases where high-Z ions are present, examining the specific density and temperature of the plasma under consideration.
In the future, these calculations could be extended to heavy and multi-charged projectiles~\cite{BraenzelPRL18, BarrigaPRE16, BarrigaPRE13}. 


%
\section{Acknowledgements}

This work was financed by the Junta de Comunidades de Castilla-La Mancha under contract No. SB-PLY/19/180501/000105, and  the Spanish Ministerio de Ciencia
e Innovación under contracts PID2019-110802GB-I00 and PID2019-110678GB-I00, and FEDER.

\bibliographystyle{elsarticle-num} 
\bibliography{stoppingplasma}

\end{document}